\documentclass[twocolumn, amsmath, amssymb, aps, notitlepage, nofootinbib,superscriptaddress]{revtex4-1}
\usepackage[normalem]{ulem}
\usepackage{bm,color,float,graphicx}
\usepackage{lipsum}
\usepackage{simplewick}
\usepackage{aas_macros}
\usepackage{natbib}

\def\t{\tilde}
\def\m{\mathrm}

\def\h{\hat}
\newcommand{\expect}[1]{\langle #1\rangle}

\date{\today}
\begin{document}
\title{Efficient estimation of rotation-induced bias to reconstructed CMB lensing power spectrum}

\author{Hongbo Cai}
\email{ketchup@sjtu.edu.cn}
\affiliation{Department of Astronomy, School of Physics and Astronomy, Shanghai Jiao Tong University, Shanghai, 200240, China}
\affiliation{Key Laboratory for Particle Astrophysics and Cosmology (MOE)/Shanghai Key Laboratory for Particle Physics and Cosmology, Shanghai, China}
\affiliation{Department of Physics and Astronomy, University of Pittsburgh, Pittsburgh, PA, USA 15260}

\author{Yilun Guan}
\email{yilun.guan@dunlap.utoronto.ca}
\affiliation{Dunlap Institute for Astronomy and Astrophysics, University of Toronto, Toronto, ON M5S 3H4, Canada}

\author{Toshiya Namikawa}
\affiliation{Center for Data-Driven Discovery, Kavli IPMU (WPI), UTIAS, The University of Tokyo, Kashiwa, 277-8583, Japan}

\author{Arthur Kosowsky}
\affiliation{Department of Physics and Astronomy, University of Pittsburgh, Pittsburgh, PA, USA 15260}

\begin{abstract}
The cosmic microwave background (CMB) lensing power spectrum is a powerful probe of the late-time universe, encoding valuable information about cosmological parameters such as the sum of neutrino masses and dark energy equation of state. However, the presence of anisotropic cosmic birefringence can bias the reconstructed CMB lensing power spectrum using CMB polarization maps, particularly at small scales, and affect the constraints on these parameters. Upcoming experiments, which will be dominated by the polarization lensing signal, are especially susceptible to this bias. We identify the dominant contribution to this bias as an $N_L^{(1)}$-like noise, caused by anisotropic rotation instead of lensing. We show that, for an CMB-S4-like experiment, a scale-invariant anisotropic rotation field with a standard deviation of 0.05 degrees can suppress the small-scale lensing power spectrum ($L\gtrsim 2000$) at a comparable level to the effect of massive neutrino with $\sum_i m_{\nu_{i}}=50~\rm{meV}$, making rotation field an important source of degeneracy in neutrino mass measurement for future CMB experiments.
We provide an analytic expression and a simulation-based estimator for this $N_L^{(1)}$-like noise, which allows for efficient forecasting and mitigation of the bias in future experiments. Furthermore, we investigate the impact of a non-scale-invariant rotation power spectrum on the reconstructed lensing power spectrum and find that an excess of power in the small-scale rotation power spectrum leads to a larger bias. Our work provides an effective numeric framework to accurately model and account for the bias caused by anisotropic rotation in future CMB lensing measurements.
\end{abstract}
\maketitle

\section{Introduction}
\label{sec:introduction}
Cosmic microwave background (CMB) lensing is an important CMB secondary effect that occurs when the trajectories of CMB photons are deflected by the gravitational effect of the large-scale structure along the paths from the last scattering surface to us. Measurements of CMB lensing probe matter distribution along the line-of-sight and is thus a powerful tool for testing cosmological models such as constraining the sum of neutrino masses and the equation of state for the dark energy.

The CMB lensing power spectrum has been measured with increasing significance by different CMB experiments, including Planck
\cite{PlanckCollaboration:2020,PlanckCollaboration:2014, carron:2022:lensing}, ACT \cite{vanEngelen:2015, ACT:2023dou}, BICEP \cite{BICEP2Collaboration:2016}, Polarbear \cite{PolarbearCollaboration:2014,POLARBEARCollaboration:2017}, and SPT
\cite{vanEngelen:2012,Story:2015}.\footnote{The CMB lensing power spectrum measurement with the state-of-the-art precision (43 $\sigma$ significance) is given by ACT \cite{ACT:2023dou}.} The upcoming experiments like Simons Observatory \cite{SO:2019:SciGoal} and CMB-S4 \cite{S4:2016:SciBook} will measure CMB lensing power spectrum with higher precision especially in the smaller angular scales, which will allow us to stringently constrain effects such as neutrino masses.

The measurement of CMB lensing power spectrum relies on the reconstruction of CMB lensing field using quadratic combination of CMB temperature and polarization maps \cite{Hu:2002,Okamoto:2003}, a technique known as the quadratic estimator. To date, temperature maps dominate the signal-to-noise in lensing detection, but expected improvements in polarization sensitivity and the fact that CMB polarization is much less affected by various contaminants, such as the thermal and kinematic Sunyaev–Zeldovich (tSZ, kSZ) effects, we expect CMB polarization maps to dominate the signal-to-noise in upcoming surveys such as CMB-S4 \cite{2016arXiv161002743A}. Therefore, it is crucial to investigate effects that may potentially bias the estimated CMB lensing power spectrum from polarization maps.

One phenomenon that may affect CMB polarization measurements is the hypothesized cosmic birefringence, which is an effect that causes the linear polarization of CMB photons to be rotated along the propagation path (see e.g. \cite{Komatsu:2022nvu} for review). Cosmic birefringence can result from the coupling between axion-like particles and photons, a coupling known as the Chern-Simons interaction \cite{Carroll:1998, Li:2008tma, Marsh:2015xka}, or from other parity-violating physics in the early universe \cite{Leon:2016kvt}, such as primordial magnetic field \cite{Kosowsky:1996yc, Harari:1996ac,Kosowsky:2004zh,2012PhRvD..86l3009Y,De:2013dra,Pogosian:2013dya}. In addition, instrumental systematics of polarization angle can also mimic this signal \cite{Mirmelstein:2020pfk, Nagata:2021}.

Cosmic birefringence may have both an isotropic and anisotropic component. Hints of isotropic cosmic birefringence have been reported in recent analyses of Planck polarization data \cite{Minami:2020, Diego-Palazuelos:2022dsq, Eskilt:2022cff}, though whether the signal is of cosmological nature is still debated. The anisotropic cosmic birefringence has also been constrained by several CMB experiments \cite{Namikawa:2020, SPT:2020cxx}, though the results are consistent with null.

Although anisotropic cosmic birefringence acts on the CMB polarization maps in an orthogonal way to that of the CMB lensing, leading to no bias to the reconstructed CMB lensing map, in a previous work \cite{Cai:2022zad} we have empirically demonstrated that the presence of anisotropic cosmic birefringence can still lead to a non-negligible bias in the reconstructed CMB lensing power spectrum, most significant on the small scales. In this paper, we expand the previous analysis in \cite{Cai:2022zad} by identifying the dominant contributions of the bias and developing a simulation-based framework to estimate the size of the bias.

The paper is organized as follows. We revisit the basics of CMB lensing and cosmic birefringence in Sec.~\ref{sec:lensing and rotation}. In Sec.~\ref{sec:CMB lensing reconstruction}, we review CMB lensing reconstruction using EB estimator and the methodology of estimating the bias from anisotropic rotation. In Sec.~\ref{sec:simulation}, we introduce the simulation procedures of $\mathcal{O}(\alpha^{2})$ contribution  and show the numerical results. Note that in this paper, for simplicity, the derivations are formulated using flat-sky approximation, while we apply full-sky simulations and CMB lensing reconstruction algorithm. We conclude in Sec.~\ref{sec:conclusion}.

\section{Lensed and Rotated CMB Power Spectra}
\label{sec:lensing and rotation}
In this section, we revisit the effects of CMB lensing and anisotropic cosmic birefringence on CMB polarization maps. We adopt flat-sky approximation throughout the paper for simplicity.



The distortion from CMB lensing can be described by lensing potential field $\phi$, and the rotation effect from anisotropic cosmic birefringence can be described by a rotation field $\alpha$. In the presence of both effects, the CMB polarization becomes
\begin{eqnarray}
  \label{eq:rot lens E}
    \t{E}'_{\bm{\ell}}&=&E_{\bm{\ell}}+\delta \t{E}_{\bm{\ell}}+ \delta E'_{\bm{\ell}} + \mathcal{O}(\phi^{n_{1}} \alpha^{n_{2}}), \\
    \label{eq:rot lens B}
   \t{B}'_{\bm{\ell}}&=&\delta \t{B}_{\bm{\ell}}+ \delta B'_{\bm{\ell}} + \mathcal{O}(\phi^{n_{1}} \alpha^{n_{2}}),
\end{eqnarray}
where $\delta \t{E}_{\bm{\ell}}$, $\delta \t{B}_{\bm{\ell}}$ refer to the first order perturbation from CMB lensing, $\delta E'_{\bm{\ell}}$, $\delta B'_{\bm{\ell}}$ denote that from the rotation field,\footnote{Note that we will follow this notational convention, denoting rotation induced quantities with prime and lensing-induced quantities with a tilde, throughout this paper.} and $\mathcal{O}(\phi^{n_{1}} \alpha^{n_{2}})$, with $n_1+n_2>1$, represents the high-order terms which mix the lensing and rotation effects. Note that we assume the primordial B-mode contribution is zero.

Specifically, for CMB lensing,
\begin{equation}
  \label{eq:1st E lens}
    \begin{aligned}
 \delta \t{E}_{\bm{\ell}} = \int \frac{d^2\bm{\ell}'}{(2 \pi)^2} &\left(\ E_{\bm{\ell}'}\cos2\varphi_{\bm{\ell'\ell}}+B_{\bm{\ell}'}\sin2\varphi_{\bm{\ell'\ell}}  \right)\\
        & \times \left[-\bm{\ell}\cdot(\bm{\ell}-\bm{\ell}') \phi_{\bm{\ell}-\bm{\ell}'}\right],
    \end{aligned}
\end{equation}

\begin{equation}
  \label{eq:1st B lens}
    \begin{aligned}
      \delta \t{B}_{\bm{\ell}} = \int \frac{d^2\bm{\ell}'}{(2 \pi)^2} &\left(-E_{\bm{\ell}'}\sin2\varphi_{\bm{\ell'\ell}}+B_{\bm{\ell}'}\cos2\varphi_{\bm{\ell'\ell}}  \right)\\
      & \times \left[-\bm{\ell}\cdot(\bm{\ell}-\bm{\ell}') \phi_{\bm{\ell}-\bm{\ell}'}\right],
    \end{aligned}
  \end{equation}
to leading-order in $\phi$, with $\varphi_{\bm{\ell'\ell}}=\varphi_{\bm{\ell'}}-\varphi_{\bm{\ell}}$. 

For anisotropic cosmic birefringence,
\begin{equation}
  \label{eq:1st E rotation}
  \begin{aligned}
  \delta E'_{\bm{\ell}} = \int \frac{d^2\bm{\ell}'}{(2 \pi)^2}
  &\left(-E_{\bm{\ell}'}\sin2\varphi_{\bm{\ell'\ell}}+B_{\bm{\ell}'}\cos2\varphi_{\bm{\ell'\ell}}  \right) \\ &\times 2\alpha_{\bm{\ell}-\bm{\ell}'},
  \end{aligned}
\end{equation}
\begin{equation}
  \label{eq:1st B rotation}
  \begin{aligned}
  \delta B'_{\bm{\ell}} = \int \frac{d^2\bm{\ell}'}{(2 \pi)^2}
  &\left(-E_{\bm{\ell}'}\cos2\varphi_{\bm{\ell'\ell}}-B_{\bm{\ell}'}\sin2\varphi_{\bm{\ell'\ell}} \right)\\
  &\times 2\alpha_{\bm{\ell}-\bm{\ell}'},
\end{aligned}
\end{equation}
to leading-order in $\alpha$. See Appendix~\ref{sec:rotation} for a review of cosmic birefringence.

Both CMB lensing and anisotropic rotation lead to off-diagonal covariance between E-mode and B-mode polarization fields \cite{Cai:2022zad}. To the leading order of $\phi$ and $\alpha$ in flat sky approximation,
\begin{equation}
  \label{eq:rotated-lensed average}
{\langle \t{E}'_{\bm{\ell}} \t{B}'_{\bm{\ell'}} \rangle}_{\m{CMB}} =  f^{\phi}_{\bm{\ell},\bm{\ell'}} \phi_{\bm{\ell}+\bm{\ell}'} + f^{\alpha}_{\bm{\ell},\bm{\ell'}} \alpha_{\bm{\ell}+\bm{\ell}'}.
\end{equation}
where $\expect{}_{\rm CMB}$ denotes the average over primary CMB with fixed lensing and rotation field realization, and 
\begin{equation}
\begin{split}
  \label{eq:f^phi}
f^{\phi}_{\bm{\ell},\bm{\ell'}} &= C^{\m{EE}}_{\ell}(\bm{\ell}+\bm{\ell}')\cdot \bm{\ell}\sin2\varphi_{\bm{\ell},\bm{\ell'}},\\
  f^{\alpha}_{\bm{\ell},\bm{\ell'}} &= 2C^{\m{EE}}_{\ell}\cos2\varphi_{\bm{\ell},\bm{\ell'}},
\end{split}
\end{equation}
which satisfies the orthogonality relation,
\begin{equation}
  \label{eq:orthogonality relation}
\int \frac{d^2 \bm{\ell}}{(2 \pi)^2} \frac{f^{\phi}_{\bm{\ell}, \bm{L}-\bm{\ell}} f_{\bm{\ell}, \bm{L}-\bm{\ell}}^{\alpha}}{\hat{C}^{\m{EE}}_{\ell} \hat{C}^{\m{BB}}_{|\bm{L}-\bm{\ell}|}} = 0.
\end{equation}
In this context, we say that CMB lensing and anisotropic cosmic birefringence are orthogonal \footnote{The full-sky orthogonality relation is given in \cite{Namikawa:2020, Cai:2022zad}} at the leading order \cite{2009PhRvD..79l3009Y}.

\section{CMB Lensing reconstruction bias due to rotation field}
\label{sec:CMB lensing reconstruction}

Taking advantage of the off-diagonal covariance induced by CMB lensing, the lensing potential can be reconstructed using a quadratic estimator \cite{Hu:2002, Okamoto:2003},
\begin{equation}
  \label{eq:estimator}
\widehat{\phi}_{\bm{L}}=\frac{A_L}{L^2} \int \frac{d^2 \bm{\ell}}{(2 \pi)^2} f^{\phi}_{\bm{\ell}, \bm{L}-\bm{\ell}}\frac{\t{E}_{\bm{\ell}}\t{B}_{\bm{L}-\bm{\ell}}}{\hat{C}^{\m{EE}}_{\ell}\hat{C}^{\m{BB}}_{|\bm{L}-\bm{\ell}|}},
\end{equation}
where $A_L$ is a normalization factor given by
\begin{equation}
  \label{}
A_L = L^2\left( \int \frac{d^2\bm{\ell}}{(2\pi)^2} \frac{|f^\phi_{\bm{\ell}, \bm{L}-\bm{\ell}} |^2}{\hat{C}^{\m{EE}}_{\ell}\hat{C}^{\m{BB}}_{|\bm{L}-\bm{\ell}|}}\right)^{-1},
\end{equation}
and $\h{C}^{\m{EE}}_{\ell}$ and $\h{C}^{\m{BB}}_{\ell}$ are the total observed EE and BB power spectra with both contributions from the noise and the lensing effect.

The lensing power spectrum is directly relevant for cosmological parameter constraints and can be estimated by
\begin{equation}
  \label{eq:phi ps estimation}
\h{C}_{L}^{\phi \phi} =  C_{L}^{\h{\phi} \h{\phi}} - {}^{(\m{RD})}N^{(0)}_{L} - N^{(1)}_{L} - N^{(\m{MC})}_{L} - N^{(\m{FG})}_{L},
\end{equation}
where
\begin{equation}
  \label{eq:bandpower}
  C_{L}^{\h{\phi}\h{\phi}}=\frac{1}{2L+1} \sum_{|\bm{L}|=L}\left|\h{\phi}_{\bm{L}}\right|^{2}.
\end{equation}
The terms subtracted in Eq.~\eqref{eq:bandpower} are different 
types of noise biases: the realization-dependent 
Gaussian noise (RDN0) 
\cite{Namikawa_2013} ${}^{(\m{RD})}N^{(0)}_{L}$, the bias 
coming from $\mathcal{O}(\phi^{2})$ terms in the connected 
contractions in CMB four-point function 
\cite{Kesden:2003, anderes2013decomposing} denoted by $N^{(1)}_{L}$, 
the ``Monte-Carlo'' (MC) noise $N^{(\m{MC})}_{L}$ and the modeled foreground bias $N^{(\m{FG})}_{L}$ (also see Sec.~III of \cite{Cai:2022zad} for a more detailed introduction of these noise biases).


In the presence of cosmic birefringence, the reconstructed lensing potential, $\hat{\phi}_L$, remains unbiased due to the orthogonality between rotation and lensing estimator. However, the lensing power spectrum may still be biased if the rotation effect is not accounted for during lensing reconstruction \cite{Cai:2022zad}.\footnote{This occurs when performing lensing reconstruction using a set of polarization maps which have been \textit{unknowingly} rotated.}

We denote the estimator applied to the rotated-lensed quantities, $\t{E}'_{\bm \ell}$, $\t{B}'_{\bm \ell}$, as
\begin{equation}
  \label{eq:estimator rot}
\widehat{\phi}'_{\bm{L}}=\frac{A_L}{L^2} \int \frac{d^2 \bm{\ell}}{(2 \pi)^2} f^{\phi}_{\bm{\ell}, \bm{L}-\bm{\ell}}\frac{\t{E}'_{\bm{\ell}}\t{B}'_{\bm{L}-\bm{\ell}}}{\hat{C}^{\m{EE}}_{\ell}\hat{C}^{\m{BB}}_{|\bm{L}-\bm{\ell}|}},
\end{equation}
and denote its reconstructed lensing power spectrum as
\begin{equation}
  \label{eq:rot phi reconstruction average}
  \begin{aligned}
 \langle \h{C'}_{L}^{\phi \phi}\rangle = \langle C_{L}^{\h{\phi}' \h{\phi}'} \rangle  - \langle {}^{(\m{RD})}N'^{(0)}_{L} \rangle &- N^{(1)}_{L} - N^{(\m{MC})}_{L}.
   \end{aligned}
\end{equation}
The bias from rotation can be defined as $\Delta(\hat{C}_{L}^{\phi \phi})_{\mathrm{rot}} = \langle \h{C'}_{L}^{\phi \phi} \rangle - \langle \h{C}_{L}^{\phi \phi} \rangle$, and can be estimated using simulations with

\begin{equation}
    \label{eq:ps bias estimator}
    \begin{aligned}
      \Delta(\hat{C}_{L}^{\phi \phi})_{\mathrm{rot}}
      = &\langle C^{\h{\phi'}\h{\phi'}}_{L} \rangle - \langle C^{\h{\phi}\h{\phi}}_{L} \rangle\\ - &(\langle {}^{(\m{RD})}\m{N}'^{(0)}_{L}  \rangle
 - \langle {}^{(\m{RD})}\m{N}^{(0)}_{L} \rangle).
 \end{aligned}
\end{equation}
Note that in practice both $\m{N}_L^{(1)}$ and $\m{N}_L^{\rm (MC)}$ are estimated
using simulations of fiducial cosmology that includes no rotation effect;
${}^{\rm (RD)}\m{N}_L^{(0)}$, on the other hand, is realization-dependent and, thus,
is affected by rotation.

In \cite{Cai:2022zad} we have demonstrated using simulations
that anisotropic birefringence leads to a non-negligible bias on
the small-scale lensing power spectrum at percent level for CMB-S3 and CMB-S4 experiments with a rotation power spectrum amplitude well below the current constraint, and the
bias scales approximately linearly with $C_L^{\alpha\alpha}$.
This suggests that the observed bias may come from terms
which contain $\alpha^2$ contribution, e.g., terms like $\mathcal{O}(\alpha^{2})$, $\mathcal{O}(\alpha^{2}\phi^{2})$.
In this work, we show that the dominant contribution to the observed
bias comes from the $\m{N}^{(1)}_L$-like noise bias caused by rotation instead of lensing, denoted as $\m{N}_L^{(1,\alpha\alpha)}$.

Calculating power spectrum using quadratic estimator involves evaluating four-point correlation function of the form
\begin{equation}
\begin{split}\label{eq:bias main term}
&\:\:\:\:\:\:\langle E_{\bm{\ell}_1}\delta B'_{\bm{\ell}_2}E_{\bm{\ell}_3}\delta B'_{\bm{\ell}_4}\rangle \\
& \propto \langle E_{\bm{\ell}_1}(E_{\bm{\ell}'_2}\alpha_{\bm{L}})E_{\bm{\ell}_3} (E_{\bm{\ell}'_4}\alpha_{\bm{L}'})\rangle.
\end{split}
\end{equation}
Assuming $\alpha$ is uncorrelated with CMB polarization field and using Wick's theorem,
\begin{equation}
  \begin{split}\label{eq:rot contractions}
    & \langle E_{\bm{\ell}_1}(E_{\bm{\ell}'_2}\alpha_{\bm{L}})E_{\bm{\ell}_3} (E_{\bm{\ell}'_4}\alpha_{\bm{L}'})\rangle \\
    = &  {\contraction {\langle} {E}{_{\bm{\ell}_1} (E_{\bm{\ell}'_2}\alpha_{\bm{L}}) }{E}
    \contraction [2ex] {\langle E_{\bm{\ell}_1}(} {E}{_{\bm{\ell}'_2}\alpha_{\bm{L}})E_{\bm{\ell}_3}(} {E}
    \bcontraction {\langle E_{\bm{\ell}_1} (E_{\bm{\ell}'_2}}{\alpha}{_{\bm{L}}) E_{\bm{\ell}_3} (E_{\bm{\ell}'_4}}{\alpha}
    \langle E_{\bm{\ell}_1} (E_{\bm{\ell}'_2}\alpha_{\bm{L}}) E_{\bm{\ell}_3} (E_{\bm{\ell}'_4}\alpha_{\bm{L}'})\rangle} \rightarrow \mathrm{(b1)} \\
  + & {\contraction {\langle}{E}{_{\bm{\ell}_1} (E_{\bm{\ell}'_2}\alpha_{\bm{L}}) E_{\bm{\ell}_3}(}{E}
    \contraction [2ex] {\langle E_{\bm{\ell}_1}(} {E}{_{\bm{\ell}'_2}\alpha_{\bm{L}})} {E}
    \bcontraction {\langle E_{\bm{\ell}_1} (E_{\bm{\ell}'_2}}{\alpha}{_{\bm{L}}) E_{\bm{\ell}_3} (E_{\bm{\ell}'_4}}{\alpha}
    \langle E_{\bm{\ell}_1} (E_{\bm{\ell}'_2}\alpha_{\bm{L}}) E_{\bm{\ell}_3} (E_{\bm{\ell}'_4}\alpha_{\bm{L}'})\rangle} \rightarrow \mathrm{(b2)} \\
  + & {\contraction {\langle}{E}{_{\bm{\ell}_1(}}  {E}
    \contraction {\langle E_{\bm{\ell}_1}({E}{_{\bm{\ell}'_2}\alpha_{\bm{L}})}} {E}{_{\bm{\ell}_3}(}{E}
    \bcontraction {\langle E_{\bm{\ell}_1} (E_{\bm{\ell}'_2}}{\alpha}{_{\bm{L}}) E_{\bm{\ell}_3} (E_{\bm{\ell}'_4}}{\alpha}
  \langle E_{\bm{\ell}_1} (E_{\bm{\ell}'_2}\alpha_{\bm{L}}) E_{\bm{\ell}_3}  (E_{\bm{\ell}'_4}\alpha_{\bm{L}'})\rangle} \rightarrow \mathrm{(b3)},
  \end{split}
\end{equation}
where the connected lines denote pairs of field to be contracted. Among the three terms, (b1) is a disconnected contribution that vanishes as a result of the orthogonality relation between the weight functions of $\phi$ and $\alpha$. (b2) and (b3) are both connected contractions that together lead to the $\m{N}_L^{(1,\alpha\alpha)}$ noise bias. Under flat-sky approximation, this bias can be calculated analytically with
\begin{equation}
  \label{eq:N1aa}
      \begin{aligned}
        N^{(1, \alpha \alpha)}_{L} = &\frac{A^2_L}{L^4} \int \frac{d^2 \bm{\ell}}{(2\pi)^2} \frac{d^2 \bm{\ell'}}{(2\pi)^2} C^{\alpha \alpha}_{\bm{\ell'}} \frac{f^{\phi}_{\bm{\ell}, \bm{L}-\bm{\ell}}}{{\hat{C}^{\m{EE}}_{\ell} \hat{C}^{\m{BB}}_{|\bm{L}-\bm{\ell}|}}} \\
         & \times \frac{f^{\phi}_{\bm{\ell}-\bm{L}-\bm{\ell}', \bm{\ell}'-\bm{\ell}}}{{\hat{C}^{\m{EE}}_{|\bm{\ell}-\bm{L}-\bm{\ell}|} \hat{C}^{\m{BB}}_{|\bm{\ell}'-\bm{\ell}|}}}\ f^{\alpha}_{\bm{\ell},\bm{\ell}'-\bm{\ell}}f^{\alpha}_{\bm{L}-\bm{\ell}, \bm{\ell}-\bm{L}-\bm{\ell}'}.
      \end{aligned}
    \end{equation}
See Appendix~\ref{sec:appendix C} for more detailed discussion on the derivation.
We calculate this expression using a modified version of
\texttt{plancklens} \cite{2020ascl.soft10009C}. Figure~\ref{fig:N1aa_plancklens}
(left panel) compares analytic calculation with simulation results and shows a
good agreement.  

Eq.~\eqref{eq:N1aa} is useful for sensitivity forecasting, but for a
more practical calculation with either full-sky geometry or masked sky, we can calculate
$N^{(1, \alpha \alpha)}_{L}$ using a simulation-based method similar to that used
in the estimation of $N^{(1)}_{L}$ for lensing \cite{Story:2015, PlanckCollaboration:2020, ACT:2023dou} as,
\begin{equation}
  \label{eq:N1aa sim}
  \begin{aligned}
 N^{(1, \alpha \alpha)}_{L} = &\langle C_{L}^{\h{\phi}\h{\phi}}[E'^{S^1_{\alpha}}B'^{S^2_{\alpha}},E'^{S^1_{\alpha}}B'^{S^2_{\alpha}}] \\
    &+ C_{L}^{\h{\phi}\h{\phi}}[E'^{S^1_{\alpha}}B'^{S^2_{\alpha}},E'^{S^2_{\alpha}}B'^{S^1_{\alpha}}]\\
    &- C_{L}^{\h{\phi}\h{\phi}}[E'^{S^1}B'^{S^2},E'^{S^1}B'^{S^2}] \\
    &- C_{L}^{\h{\phi}\h{\phi}}[E'^{S^1}B'^{S^2},E'^{S^2}B'^{S^1}] \rangle_{S^1_{\alpha},S^2_{\alpha},S^1,S^2}.
  \end{aligned}
\end{equation}
Here $(S^1_{\alpha},S^2_{\alpha})$ labels pairs of rotated CMB maps with the same realization of rotation field $\alpha$ but different realizations of primary CMB; $(S^1,S^2)$ labels pairs of rotated CMB maps with different realizations of both the primary CMB and $\alpha$. Eq.~\eqref{eq:N1aa sim} remains valid in full-sky or masked sky and hence will be used to compare with the observed bias in the next section. In Sec.~\ref{sec:simulation}, by computing $N^{(1, \alpha \alpha)}_{L}$ using both the simulation-based and analytic approaches, we identify $\mathcal{O}(\alpha^{2})$ as the dominant bias contribution.

\begin{figure*}[t]
\includegraphics[width=0.8\textwidth]{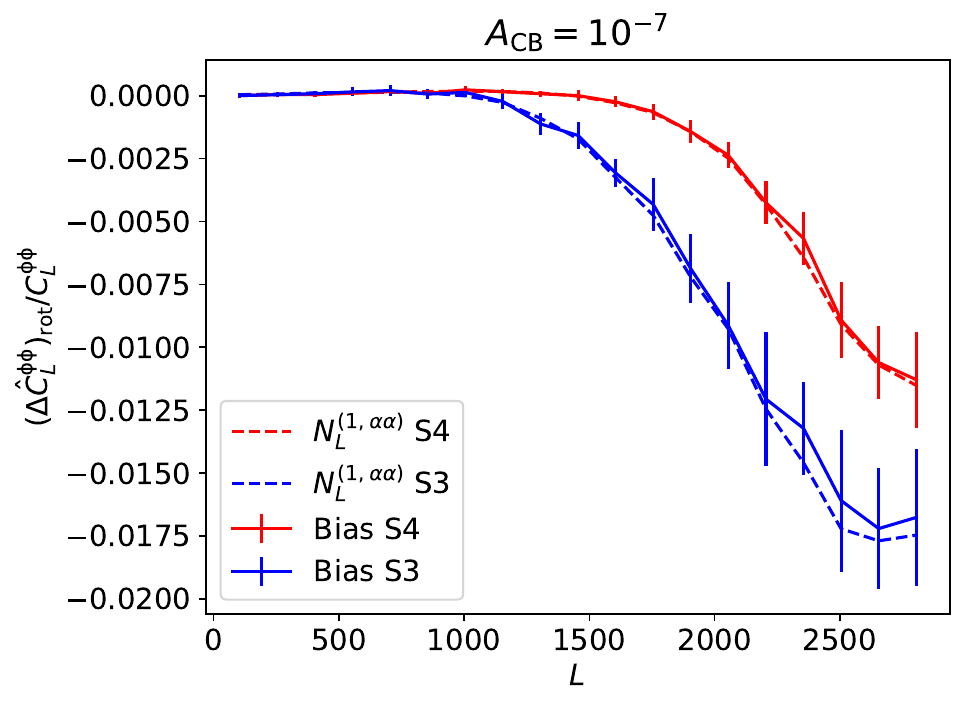}
\caption{The comparison of the simulated $N^{(1, \alpha \alpha)}_{L}/C_{L}^{\phi \phi}$ (solid lines) and the total bias of reconstructed lensing potential power spectrum $\Delta(\hat{C}_{L}^{\phi \phi})_{\mathrm{rot}}$ (dashed lines) from a scale-invariant anisotropic rotation field with amplitude of $A_{\m{CB}}=10^{-7}$ for CMB-S3-like experiment and CMB-S4-like experiment. We show the fractional values with a factor of $1/C_{L}^{\phi \phi}$ which is the CMB lensing power spectrum from the fudical model. We apply full-sky CMB lensing reconstruction using the EB estimator, and the CMB multipole ranges from $\ell_{\mathrm{min}} = 30$ to $\ell_{\mathrm{max}}=3000$, and the curves are binned by $\Delta \ell =150$.}
    \label{fig:comparison}
    \centering
  \end{figure*}

\begin{figure*}[t]
\includegraphics[width=0.8\textwidth]{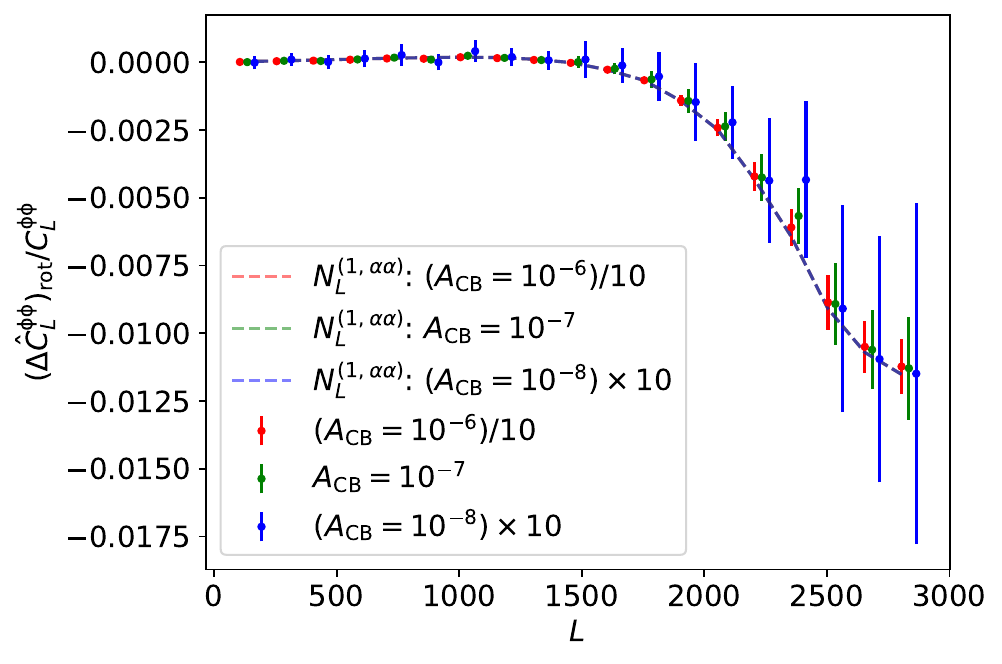}
    \caption{The fractional $N^{(1, \alpha \alpha)}_{L}/C_{L}^{\phi \phi}$ (dashed lines) and the fractional bias $\Delta(\hat{C}_{L}^{\phi \phi})_{\mathrm{rot}}/C_{L}^{\phi \phi}$ (solid lines) with $A_{\m{CB}}=10^{-6}, 10^{-7}$ and $10^{-8} $ for a CMB-S4-like experiment. For visualization we have scaled both data and error bars by a factor of
10 and 0.1 for $A_{\rm CB}=10^{-8}$ and $10^{-6}$, respectively. The horizontal displacements between the three data series are for visualization purpose only.}
    \label{fig:scaling}
    \centering
\end{figure*}

\section{Simulation and Results}
\label{sec:simulation}
In this section, we calculate $N^{(1, \alpha \alpha)}_{L}$
using simulation-based method following Eq.~\eqref{eq:N1aa sim}.
We then compare it with the total bias,
$\Delta(\hat{C}_{L}^{\phi \phi})_{\mathrm{rot}}$,
from \cite{Cai:2022zad}. All simulations in this work are
generated in CAR pixelization on the full sky
using \texttt{pixell}
\footnote{\url{https://github.com/simonsobs/pixell}}\cite{2021ascl.soft02003N}.
We use \texttt{cmblensplus}\footnote{\url{https://github.com/toshiyan/cmblensplus}} to perform the CMB lensing reconstruction on the full sky which avoids unnecessary complications due to partial sky coverage.\footnote{There can be extra mean-field bias for cut-sky CMB lensing reconstruction \cite{Namikawa_2013}.}
Nonetheless Eq.~\eqref{eq:N1aa sim} is applicable to both full-sky maps and partial-sky maps. The simulation code is publicly available on \texttt{github}.\footnote{\url{https://github.com/catketchup/lens_rot_bias}}

\begin{table}[tp]
\centering
\begin{tabular}{|p{2.4cm}|p{1.4cm}|p{1.4cm}|}
 \hline
 Expt & ${\Delta}_{T}~[\mathrm{\mu K '}]$ & ${\theta}_{\mathrm{FWHM}} [']$\\[0.5ex]
  \hline
  CMB-S3-like & 7 & 1.4 \\
  CMB-S4-like & 1 & 1.4 \\
 \hline
\end{tabular}
\caption{Experiments configurations considered in this study same as the one used in \cite{Cai:2022zad}.}
\label{tab:expts}
\end{table}

Note that though sharing similar simulation technique and experimental configuration, \cite{Cai:2022zad} estimates the rotation-induced bias directly whereas this work simulates $N^{(1, \alpha \alpha)}_{L}$ to identify if $\mathcal{O}(\alpha^{2})$ contribution is dominant. The latter one avoids the computationally-expensive lensed simulations and RDN0 calculation. 

To estimate $\m{N}_L^{(1,\alpha\alpha)}$ using Eq.~\eqref{eq:N1aa sim}, we generate simulations with the following steps:
\begin{enumerate}
\item
We generate two sets of 10 realizations of full-sky primary
CMB polarization maps (unlensed) based on a fiducial power
spectrum that does not include the cosmic birefringence effect
from the best-fit cosmology from Planck 2018 \cite{Planck2018:VI:CP}.
\item
We generate 11 realizations of
anisotropic rotation field. We assume the rotation field, $\alpha$, follows Gaussian statistics with a scale-invariant rotation power spectrum, $C_L^{\alpha\alpha}$, given by
\begin{equation}
  \label{eq:claa}
\frac{L(L+1) C_{L}^{\alpha \alpha}}{2 \pi}\equiv A_{\mathrm{CB}},
\end{equation}
where $A_{\m{CB}}$ characterizes the amplitude of the rotation power spectrum, with $A_{\m{CB}}=10^{-7}$ being the expected $1\sigma$ upper-limit for CMB-S3-like experiments \cite{2019PhRvD.100b3507P} corresponding to an rms rotation of around 0.05 degrees.
\item
We perform a pixel-wise polarization rotation on the first set of
10 primary CMB polarization maps with the
first realization of rotation field to make a set of 10 rotated
CMB polarization maps $S_{\m{\alpha}}$.
\item
We then rotate each realization in the second set of ten primary
CMB polarization maps using each one of the remaining 10 rotation
field realizations to make the second set of 10 rotated CMB
polarization maps labeled by $S$.
\item
We estimate $N^{(1, \alpha \alpha)}_{L}$ using
Eq.~\eqref{eq:N1aa sim}, where $(S^1_{\alpha},S^2_{\alpha})$ and $(S^1,S^2)$ are two pairs of simulations drawn from the set of $S_{\alpha}$ and $S$, respectively. A total of 90 unique pairs are drawn for $S$ and $S_\alpha$. We then perform full-sky CMB lensing reconstruction
using \texttt{cmblensplus}\footnote{\url{https://github.com/toshiyan/cmblensplus}}
with CMB multipoles between $\ell_{\m{min}} = 30$ and $\ell_{\m{max}} = 3000$ for lensing reconstruction and
estimate the total power spectrum,
$\hat{C}_\ell^{XX}$ ($X\in \{E, B\}$), using the Knox
formula \cite{PhysRevD.52.4307},
\begin{equation}
  \hat{C}_\ell^{XX} = C_\ell^{XX}\vert_{\rm fid} + N_\ell,
\end{equation}
where the fiducial power spectrum, $C_\ell^{XX}\vert_{\rm fid}$, is given by the best-fit cosmology from Planck 2018 \cite{Planck2018:VI:CP} without cosmic birefringence, and the noise power spectrum for polarization is given by
\begin{equation}
  N_\ell = \Delta^2_{\m{P}} e^{\ell(\ell+1) \theta^2_{\m{FWHM}}/(8\ln2)},
\end{equation}
where $\Delta_{\m{P}}$ is the polarization noise level of the experiment, and $\theta_{\m{FWHM}}$ is the full-width at half maximum (FWHM) of the beam in radians. We consider several experimental configurations as listed in Table~\ref{tab:expts} following \cite{Cai:2022zad}.

\item We repeat the above procedures with $A_{\m{CB}}=10^{-6}, 10^{-7}$ and $10^{-8}$ for CMB-S3-like and CMB-S4-like experiments.

\end{enumerate}

In Fig.~\ref{fig:comparison}, we compare the simulated $N^{(1, \alpha \alpha)}_{L}$ with the total bias, $\Delta(\hat{C}_{L}^{\phi \phi})_{\mathrm{rot}}$, for $A_{\m{CB}}=10^{-7}$ and for a CMB-S3-like and a CMB-S4-like experiment respectively. We show the fractional bias relative to the fiducial lensing power spectrum, $C_{L}^{\phi \phi}$. Both $\Delta(\hat{C}_{L}^{\phi \phi})_{\mathrm{rot}}$ and its error bar are taken directly from \cite{Cai:2022zad}. One can see that $N^{(1, \alpha \alpha)}_{L}$ closely aligns with the total observed bias $\Delta(\hat{C}_{L}^{\phi \phi})_{\mathrm{rot}}$ for both experimental configurations.

In Fig.~\ref{fig:scaling}, we show $N^{(1, \alpha \alpha)}_{L}$ for different $A_{\rm CB}$. For visualization we have scaled both data and error bar by a factor of 10 and 0.1 for $A_{\rm CB}=10^{-8}$ and $10^{-6}$, respectively. The results show that $N^{(1, \alpha \alpha)}_{L}$ scales linearly with $A_{\m{CB}}$ and matches well with the observed behavior of the total bias for different $A_{\m{CB}}$.

Our results confirm that $N^{(1, \alpha \alpha)}_{L}$ is the dominant contribution to the rotation-induced bias in the lensing power spectrum and naturally explain its linear scaling with $A_{\mathrm{CB}}$; higher order corrections contribute only sub-dominantly and are likely the source of the slight over-estimation of the bias noticeable in Fig.~\ref{fig:comparison} though the level is negligible compared to the error bars.

Massive neutrino is also known to suppress the CMB lensing power spectrum at the small scales ($L \gtrsim 1000$) at a percent level for $\sum_i m_{\nu_i} \approx 50$~meV)\cite{TopicalConvenersKNAbazajianJECarlstromATLee:2013bxd}. Bias from rotation field may therefore become a source of degeneracy to massive neutrino. Our results suggest that a rotation field with $A_{\rm CB}=10^{-7}$ (corresponding to an rms rotation of
around 0.05 degrees) may suppress the lensing power spectrum by around 1\% at $L\gtrsim2000$ for a CMB-S4-like experiment, which is comparable to the suppression from massive neutrino with $\sum_i m_{\nu_i} \approx 50~{\rm meV}$ at these scales. This effect can be mitigated by separately fitting the rotation field and then subtract the expected bias from the lensing power spectrum. Effective mitigation of rotation-induced bias will be crucial for measuring of neutrino mass in the upcoming experiments. We leave a detailed analysis of mitigation strategy future work.

It is also worth noting that rotation-induced bias occurs only when polarization maps are used in lensing reconstruction. Lensing power spectrum measured from the TT estimator will thus differ from that from the EB estimator in the small scales. One can in theory use this difference reveal this bias, but we expect this to be roughly one order of magnitude less sensitive than directly reconstructing the rotation field, with $\sigma(A_{\rm CB}) \gtrsim 10^{-6}$. Nonetheless the difference can still be used for consistency check when such a rotation field is detected. Similar effects can be seen across other estimators such as EE and TE but are expected to yield lower signal to noise. We leave a detailed analysis of the effect on different quadratic estimators to a future work.

Note that although in this paper we choose to focus on the effect of a rotation field on the CMB lensing power spectrum, the reverse is also true that the estimated rotation power spectrum will also biased by CMB lensing due to the same mechanism \cite{Naokawa:2023upt, Liu:2016dcg}. Such effect has previously been analyzed in \cite{Bianchini:2020, BICEPKeck:2022kci} and found to be subdominant under the current sensitivity. We leave a more detailed analysis for this reverse effect in a future work.

Our results suggest that $N^{(1, \alpha \alpha)}_{L}$ can be used as an efficient way to evaluate the expected bias, $\Delta(\hat{C}_{L}^{\phi \phi})_{\mathrm{rot}}$, caused by rotation. This is important as computing $\Delta(\hat{C}_{L}^{\phi \phi})_{\mathrm{rot}}$ through $N^{(1, \alpha \alpha)}_{L}$ only uses a small number of pixel-wise rotation simulations and avoids generating an order of magnitude more computationally costly lensed simulations required to estimate RDN0 accurately for each mock data as done in \cite{Cai:2022zad}; this method also reduces the total number of lensing reconstruction needed by an order of magnitude, which is more computationally efficient.

\begin{figure*}[t]
\includegraphics[width=1\textwidth]{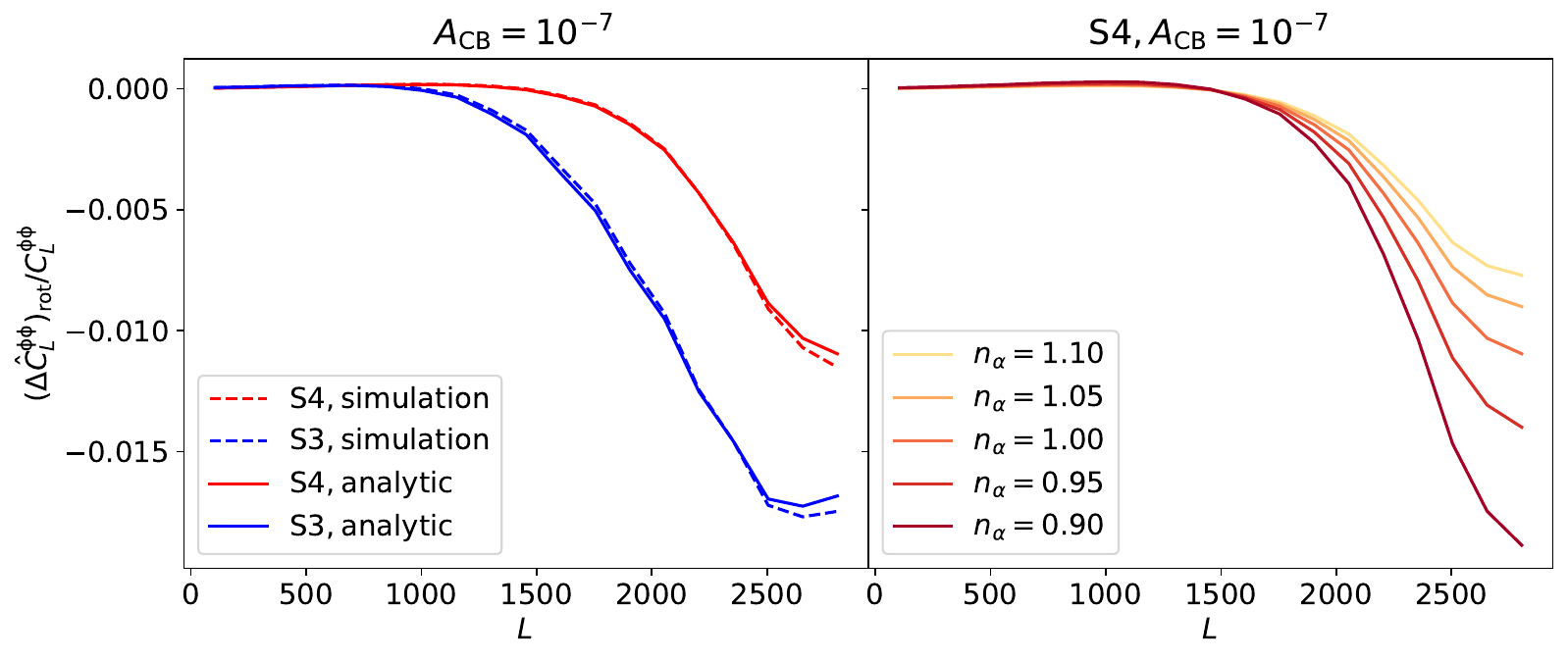}
\caption{The left panel shows the full-sky fractional bias $N^{(1, \alpha \alpha)}_{L}/C_{L}^{\phi \phi}$  from simulation (solid lines, labeled by \textit{simulation}) and from the analytic calculation (dashed lines, labeled by \textit{analytic}) using a modified version of \texttt{plancklens} for CMB-S3-like and CMB-S4-like experiments with $A_{\rm CB}=10^{-7}$. In the right panel, for a CMB-S4-like experiment, we show the analytic $N^{(1, \alpha \alpha)}_{L}$ with  scale-dependent rotation power spectrum of $C_L^{\alpha\alpha} =  A_{\rm CB}/(L(L+1))^{n_\alpha}$ for $n_{\alpha}=\{1.10, 1.05, 1.0, 0.95, 0.90\}$ and $A_{\rm CB}=10^{-7}$. }
\label{fig:N1aa_plancklens}
\centering
\end{figure*}

As a verification, we also implement a calculator for Eq.~\eqref{eq:N1aa} by modifying \texttt{plancklens} \cite{2020ascl.soft10009C} and compare the analytic calculation with the simulation results as shown in the left panel of Figure~\ref{fig:N1aa_plancklens}. We note the good agreement between the analytic calculation and simulations. The minor discrepancy in the small angular scales is likely due to the inaccuracy of the flat-sky approximation assumed in our analytic calculator.

Making use of the analytic calculator, we investigate a scenario where the rotation power spectrum deviates from a scale-invariant power spectrum. Such a scenario can occur in physical models like domain walls \cite{Kitajima:2022jzz}. To see the effect of scale-dependent rotation power spectrum on the lensing power spectrum, we parameterize the rotational power spectrum by $C_L^{\alpha\alpha} =  A_{\rm CB}/(L(L+1))^{n_\alpha}$, where $n_\alpha$ characterizes the scale dependence, with $n_\alpha=1$ being the scale-invariant case, $n_\alpha > 1$ being a red spectrum, and $n_\alpha < 1$ being a blue spectrum. Using the modified \texttt{plancklens}, we calculate the corresponding analytic $N^{(1, \alpha \alpha)}_{L}$ with $n_\alpha=\{1.10, 1.05, 1.0, 0.95, 0.90\}$, respectively, and show the results in the right panel of Fig.~\ref{fig:N1aa_plancklens}. It shows clearly that a bluer rotation power spectrum leads to a larger bias, with $n_\alpha=0.9$ leading to a 50\% larger bias in the small scales, compared to the scale-invariant case. This suggests that an excess of rotation on the small-scale, as suggested by models such as domain wall, may lead to a more significant bias in the reconstructed lensing power spectrum, making it an important effect to consider when measuring neutrino mass using lensing power spectrum.

\section{Conclusion}
\label{sec:conclusion}
The CMB lensing power spectrum, measuring the matter fluctuations over a broad range of redshifts, encodes a wealth of information of the late-time universe. Bias to CMB lensing power spectrum especially at small scales can lead to bias in constraining cosmological parameters extracted from it such as the sum of neutrino masses and $\sigma_{8} \Omega_{\m{m}}^{0.25}$. As the polarization lensing signal is going to dominate the signal-to-noise for the upcoming experiments, it is necessary to check the potential bias to the polarization-based CMB lensing power spectrum reconstruction.

Ref.~\cite{Cai:2022zad} has demonstrated that the existence of an anisotropic rotation with an amplitude as low as $A_{\rm CB}=10^{-7}$ may lead to a percent-level bias in the reconstructed CMB lensing power spectrum for the upcoming CMB experiments such as Simons Observatory and CMB-S4, and this bias scales linearly with $A_{\rm CB}$. In this work we identify the dominant contribution to the observed bias as an N1-like noise caused by rotation instead of lensing, named $N^{(1, \alpha \alpha)}_{L}$ from $\mathcal{O}(\alpha^2)$ contribution. We provide an analytic expression for this term and demonstrate a simulation-based method to estimate it similar to that used for $N^{(1)}_{L}$ in CMB lensing. We show that both the simulation-based method and direct analytic calculation of $N^{(1, \alpha \alpha)}_{L}$ agree with the observed bias in the lensing power spectrum in \cite{Cai:2022zad}.

The analytic expression provides an efficient calculator for forecasting the size of this effect for future experiments. Our calculation showed that, for a CMB-S4-like experiment with an unknown rotation field at the level of $A_{\rm CB} = 10^{-7}$, the size of the bias in CMB lensing power spectrum is at percent level in small scales ($L \gtrsim 2000$). We expect neutrino mass to affect the CMB lensing power spectrum at similar level and scales \cite{TopicalConvenersKNAbazajianJECarlstromATLee:2013bxd}.
The estimation of anisotropic cosmic birefringence and the subtraction of the rotation-induced bias from CMB lensing power spectrum will be a key degeneracy mitigation step for the measurement of neutrino mass for future CMB experiments.

With the analytic calculator we further explored the effect of a scale-dependent rotation power spectrum on the reconstructed lensing power spectrum, and we found an enhanced bias by a factor of a few when the rotation field has an excess of power in the small scales. To properly account for this bias, we also describe a simulation-based method which provides a computationally efficient way to estimate the size of this bias in a realistic experiment. These methods will allow future experiments to accurately model and mitigate the bias caused by rotation either from physics beyond the standard model such as cosmic birefringence or instrument systematics, which will be crucial to the future measurement of neutrino mass using CMB lensing.

\section*{Acknowledgments}
We thank Pengjie Zhang, Antony Lewis, Julien Carron and Omar
Darwish for very helpful discussion. HC acknowledges support from
the National Key R\&D Program of China (2023YFA1607800, 2023YFA1607801, 2020YFC2201602),
the National Science Foundation of China (11621303),
CMS-CSST-2021-A02, and the Fundamental Research Funds for
the Central Universities. This work uses resources of the
National Energy Research Scientific Computing Center (NERSC),
and open source software including
\texttt{healpy} \cite{2019JOSS....4.1298Z},
\texttt{pixell} \cite{2021ascl.soft02003N}, \texttt{cmblensplus}\cite{2021ascl.soft04021N} and \texttt{plancklens} \cite{2020ascl.soft10009C}.
TN acknowledges support from JSPS KAKENHI Grant No. JP20H05859 and No. JP22K03682, and World Premier International Research Center Initiative (WPI Initiative), MEXT, Japan.

\appendix

\section{Cosmic Birefringence}
\label{sec:rotation}
In this section, we review cosmic birefringence which refers to rotations that CMB linear polarization field may encounter from the last scattering surface to our observation point. The occurrence of cosmic birefringence can be  attributed potentially to parity-violating phenomena in the early universe, such as the interaction between axionlike particles and photons through the Chern-Simons mechanism, \cite{Carroll:1998, Li:2008tma, Marsh:2015xka}, more general Lorentz-violating physics beyond the Standard Model \cite{Leon:2016kvt}, and primordial magnetic fields through Faraday rotation which is frequency-dependent \cite{Kosowsky:1996yc, Harari:1996ac,Kosowsky:2004zh,2012PhRvD..86l3009Y,De:2013dra,Pogosian:2013dya}.

In the presence of cosmic birefringence, the rotated CMB linear polarization field can be expressed as
\begin{equation}
\label{eq:rotation}
(Q'\pm iU')(\hat{\bm{n}}) = e^{\pm 2i\alpha(\hat{\bm{n}})}(Q \pm i U)(\hat{\bm{n}}),
\end{equation}
with $\alpha(\hat{\bm{n}})$ a rotation field caused by cosmic birefringence, and the rotated quantities are labeled with prime. As usually done in the literature, a generic rotation field is split into an isotropic and an anisotropic part as
\begin{equation}
  \label{}
\alpha(\hat{\bm{n}})=\bar{\alpha}+\delta \alpha(\hat{\bm{n}}),
\end{equation}
with $\bar{\alpha}$ the isotropic part, and $\delta \alpha(\hat{\bm{n}})$ the anisotropic part with a zero mean
\begin{equation}
  \label{eq:rotation parts}
  \expect{\delta \alpha(\hat{\bm{n}})}=0.
\end{equation}
Following \cite{Cai:2022zad}, we consider cosmic birefringence induced by a Chern-Simons-type interaction between axionlike particles and photons, an well-motivated extension of the standard model which has been studied in a number of previous papers \cite{1997PhRvD..55.6760C, 1998PhRvD..58k6002C,Leon:2017} with a Lagrangian given by
\begin{equation}
  \label{eq:C-S interaction}
  \mathcal{L}_{c s}= \frac{g_{a\gamma}a}{4} F^{\mu \nu} \tilde{F}_{\mu \nu},
\end{equation}
where $a$ is the pseudoscalar field for the axionlike particle, $F^{\mu \nu}$ is the electromagnetic tensor with $\tilde{F}_{\mu \nu}$ being its dual, and $g_{a\gamma}$ is a coupling constant with a dimension of energy. The rotation angle can be given by
\begin{equation}
  \label{eq:alpha and phi}
  \alpha(\hat{\bm{n}})=\frac{g_{a\gamma}}{2}\Delta a,
\end{equation}
where $\Delta a$ is the change of $a$ along the photon path from the last scattering surface to us. Following \cite{Cai:2022zad}, in this paper we choose to focus on the anisotropic cosmic birefringence, i.e., $\bar{\alpha}=0$.

Note that depending on models, axionlike particle can be massive
or massless. Especially, if the axionlike field is
inflation-seeded and massless with a potential that vanishes,
the cosmic birefringence power spectrum at large scales
($L\lesssim 100$) is approximately
scale-invariant \cite{2011PhRvD..84d3504C}
\footnote{For massive axionlike particle,
see, e.g.,\cite{Passaglia:2022bcr, Leon:2017}}. In this
scenario, the power spectrum is connected to the inflationary
Hubble parameter $H_{\m{I}}$ and can be given by
\begin{equation}
  \label{eq:claa}
\frac{L(L+1) C_{L}^{\alpha \alpha}}{2 \pi}=\left(\frac{H_{\m{I}} g_{a \gamma}}{4 \pi}\right)^{2}\equiv A_{\mathrm{CB}},
\end{equation}
where $A_{\rm CB}$ is a dimensionless parameter to parametrize
the amplitude of the scale-invariant rotation power spectrum.
The current tightest constraint of the scale-invariant
rotation power spectrum is given by ACTPol \cite{Namikawa:2020}
and SPTPol \cite{SPT:2020cxx} corresponding to a $2\sigma$
upper limit \footnote{Note that $A_{\rm{CB}}$ defined in this
paper is $10^{-4}$ times of that in \cite{Namikawa:2020}
and \cite{SPT:2020cxx}.} of $A_{\m{CB}} \leq 10^{-5}$.
The next-generation ground-based CMB experiments expect to give
the constraint of $A_{\rm CB}$ at the level of $10^{-7}$
\cite{CMB-HD:2022bsz, Pogosian:2019jbt, Mandal:2022tqu}.

Although it is often assumed that anisotropic birefringence
follows a scale-invariant power spectrum, some physical models
are known to produce scale-dependent birefringence such as
domain walls \cite{Kitajima:2022jzz}. We discuss the effects
of scale dependence in Section~\ref{sec:simulation}.


\section{$N^{(1, \alpha \alpha)}_{L}$ Calculation}
\label{sec:appendix C}
Here we derive
$N^{(1, \alpha \alpha)}_{L}$
under the flat-sky approximation. As a comparison, we start by reviewing $N^{(1)}_{L}$ calculation for CMB lensing.
Calculating $\langle C_{L}^{\h{\phi}' \h{\phi}'} \rangle$ involves the four-point correlation function,
\begin{equation}
  \label{eq:trispec rot lens}
  \langle \t{E}'_{\bm{\ell}_1} \t{B}'_{\bm{\ell}_2} \t{E}'_{\bm{\ell}_3} \t{B}'_{\bm{\ell}_4}\rangle,
\end{equation}
which picks up an $\mathcal{O}(\phi^{2})$ contribution as
\begin{equation}
\begin{split}\label{eq:phi bias main term}
&\:\:\:\:\:\:\langle E_{\bm{\ell}_1}\delta \t{B}_{\bm{\ell}_2}E_{\bm{\ell}_3}\delta \t{B}_{\bm{\ell}_4}\rangle \\
& \propto \langle E_{\bm{\ell}_1}(E_{\bm{\ell}'_2}\phi_{\bm{L}})E_{\bm{\ell}_3} (E_{\bm{\ell}'_4}\phi_{\bm{L}'})\rangle
\end{split}
\end{equation}
where we have applied Eq.~\eqref{eq:1st B lens} and used parentheses to indicate groupings of terms which will be convenient later on. We assume $\phi$ to be a Gaussian random field and is uncorrelated with either the primary CMB or the rotation field, $\alpha$. The second line of Eq.~\eqref{eq:phi bias main term} can be expanded using Wick theorem into products of two-point correlation functions as three types,
\begin{equation}
  \begin{split}\label{eq:bias main term}
    & \langle E_{\bm{\ell}_1}(E_{\bm{\ell}'_2}\phi_{\bm{L}})E_{\bm{\ell}_3} (E_{\bm{\ell}'_4}\phi_{\bm{L}'})\rangle \\
    = &  {\contraction {\langle} {E}{_{\bm{\ell}_1} (E_{\bm{\ell}'_2}\phi_{\bm{L}}) }{E}
    \contraction [2ex] {\langle E_{\bm{\ell}_1}(} {E}{_{\bm{\ell}'_2}\phi_{\bm{L}})E_{\bm{\ell}_3}(} {E}
    \bcontraction {\langle E_{\bm{\ell}_1} (E_{\bm{\ell}'_2}}{\phi}{_{\bm{L}}) E_{\bm{\ell}_3} (E_{\bm{\ell}'_4}}{\alpha}
    \langle E_{\bm{\ell}_1} (E_{\bm{\ell}'_2}\phi_{\bm{L}}) E_{\bm{\ell}_3} (E_{\bm{\ell}'_4}\phi_{\bm{L}'})\rangle} \rightarrow \mathrm{(a1)} \\
  + &  {\contraction {\langle}{E}{_{\bm{\ell}_1} (E_{\bm{\ell}'_2}\phi_{\bm{L}}) E_{\bm{\ell}_3}(}{E}
    \contraction [2ex] {\langle E_{\bm{\ell}_1}(} {E}{_{\bm{\ell}'_2}\phi_{\bm{L}})} {E}
    \bcontraction {\langle E_{\bm{\ell}_1} (E_{\bm{\ell}'_2}}{\phi}{_{\bm{L}}) E_{\bm{\ell}_3} (E_{\bm{\ell}'_4}}{\phi}
    \langle E_{\bm{\ell}_1} (E_{\bm{\ell}'_2}\phi_{\bm{L}}) E_{\bm{\ell}_3} (E_{\bm{\ell}'_4}\phi_{\bm{L}'})\rangle} \rightarrow \mathrm{(a2)}\\
  + &  {\contraction {\langle}{E}{_{\bm{\ell}_1(}}  {E}
    \contraction {\langle E_{\bm{\ell}_1}({E}{_{\bm{\ell}'_2}\phi_{\bm{L}})}} {E}{_{\bm{\ell}_3}(}{E}
    \bcontraction {\langle E_{\bm{\ell}_1} (E_{\bm{\ell}'_2}}{\phi}{_{\bm{L}}) E_{\bm{\ell}_3} (E_{\bm{\ell}'_4}}{\phi}
  \langle E_{\bm{\ell}_1} (E_{\bm{\ell}'_2}\phi_{\bm{L}}) E_{\bm{\ell}_3} (E_{\bm{\ell}'_4}\phi_{\bm{L}'})\rangle} \rightarrow \mathrm{(a3)},
  \end{split}
\end{equation}
where $(\mathrm{a1})$ is a disconnected term, and $(\mathrm{a2})$ and $(\mathrm{a3})$ are connected terms of Eq.~\eqref{eq:phi bias main term}. It is easy to see \cite{Cooray:2002py,Kesden:2003,Jenkins:2014hza} that the contribution of $(\mathrm{a1})$ in $\langle C^{\h{\phi'}\h{\phi'}}_{L} \rangle$ is given by
\begin{equation}
  \label{eq:clpp term}
        \begin{aligned}
          C_L^{\phi \phi}\left(\frac{A_{\bm{L}}}{L^2}\right)^2 &\int \frac{d^2 \bm{\ell}}{(2 \pi)^2} \frac{|f^{\phi}_{\bm{\ell}, \bm{L}-\bm{\ell}}|^2}{{\hat{C}^{\m{EE}}_{\ell} \hat{C}^{\m{BB}}_{|\bm{L}-\bm{\ell}|}}} \\
          \times &\int \frac{d^2 \bm{\ell}'}{(2 \pi)^2} \frac{|f_{\bm{\ell}', \bm{L}-\bm{\ell}'}^{\phi}|^2}{{\hat{C}^{\m{EE}}_{\ell'} \hat{C}^{\m{BB}}_{|\bm{L}-\bm{\ell}'|}}}=C_L^{\phi \phi},
      \end{aligned}
\end{equation}
and $(\mathrm{a2})$ and $(\mathrm{a3})$ together generate the $N^{(1)}_{L}$ bias as
\begin{equation}
  \label{eq:N1}
      \begin{aligned}
        N^{(1)}_{L} = &\frac{A^2_L}{L^4} \int \frac{d^2 \bm{\ell}}{(2\pi)^2} \frac{d^2 \bm{\ell'}}{(2\pi)^2} C^{\phi \phi}_{\bm{\ell'}} \frac{f^{\phi}_{\bm{\ell}, \bm{L}-\bm{\ell}}}{{\hat{C}^{\m{EE}}_{\ell} \hat{C}^{\m{BB}}_{|\bm{L}-\bm{\ell}|}}} \\
         & \times \frac{f^{\phi}_{\bm{\ell}-\bm{L}-\bm{\ell}', \bm{\ell}'-\bm{\ell}}}{{\hat{C}^{\m{EE}}_{|\bm{\ell}-\bm{L}-\bm{\ell}|} \hat{C}^{\m{BB}}_{|\bm{\ell}'-\bm{\ell}|}}}\ f^{\phi}_{\bm{\ell},\bm{\ell}'-\bm{\ell}}f^{\phi}_{\bm{L}-\bm{\ell}, \bm{\ell}-\bm{L}-\bm{\ell}'}.
      \end{aligned}
    \end{equation}

$N^{(1)}_L$ bias can be estimated using pairs of noiseless lensed CMB maps \cite{Story:2015}. For EB estimator,
\begin{equation}
  \label{eq:N1 sim}
  \begin{aligned}
 N^{(1)}_{L} = &\langle C_{L}^{\h{\phi}\h{\phi}}[\t{E}^{S^{1}_{\phi}}\t{B}^{S^{2}_{\phi}},\t{E}^{S^{1}_{\phi}}\t{B}^{S^{2}_{\phi}}] \\
    &+ C_{L}^{\h{\phi}\h{\phi}}[\t{E}^{S^{1}_{\phi}}\t{B}^{S^{2}_{\phi}},\t{E}^{S^{2}_{\phi}}\t{B}^{S^{1}_{\phi}}]\\
    &- C_{L}^{\h{\phi}\h{\phi}}[\t{E}^{S^1}\t{B}^{S^2},\t{E}^{S^1}\t{B}^{S^2}] \\
    &- C_{L}^{\h{\phi}\h{\phi}}[\t{E}^{S^1}\t{B}^{S^2},\t{E}^{S^2}\t{B}^{S^1}] \rangle_{S^{1}_{\phi},S^{2}_{\phi},S^1,S^2},\\
  \end{aligned}
\end{equation}
where $(S^{1}_{\phi},S^{2}_{\phi})$ labels lensed CMB simulations with different primary CMB realizations but the same lensing potential realization, and $(S^1,S^2)$ labels lensed CMB simulations with different realizations for both the primary CMB and the lensing potential. This simulation method works both for full-sky geometry and masked sky and has been applied in previous precision lensing measurements such as \cite{PlanckCollaboration:2020, ACT:2023dou}.

In addition to the $\mathcal{O}(\phi^{2})$ term in Eq.~\eqref{eq:phi bias main term}, Eq.~\eqref{eq:trispec rot lens} also contains an $\mathcal{O}(\alpha^{2})$ contribution given by
\begin{equation}
\begin{split}\label{eq:bias main term}
&\:\:\:\:\:\:\langle E_{\bm{\ell}_1}\delta B'_{\bm{\ell}_2}E_{\bm{\ell}_3}\delta B'_{\bm{\ell}_4}\rangle \\
& \propto \langle E_{\bm{\ell}_1}(E_{\bm{\ell}'_2}\alpha_{\bm{L}})E_{\bm{\ell}_3} (E_{\bm{\ell}'_4}\alpha_{\bm{L}'})\rangle,
\end{split}
\end{equation}
and this leads to a $N^{(1)}$-like bias caused by rotation similar to lensing which we call $N^{(1,\alpha\alpha)}$.
Assuming that $\alpha$ is a Gaussian random field that is  uncorrelated with the primary CMB and using Wick theorem, we have
\begin{equation}
  \begin{split}\label{eq:rot contractions}
    & \langle E_{\bm{\ell}_1}(E_{\bm{\ell}'_2}\alpha_{\bm{L}})E_{\bm{\ell}_3} (E_{\bm{\ell}'_4}\alpha_{\bm{L}'})\rangle \\
    = &  {\contraction {\langle} {E}{_{\bm{\ell}_1} (E_{\bm{\ell}'_2}\alpha_{\bm{L}}) }{E}
    \contraction [2ex] {\langle E_{\bm{\ell}_1}(} {E}{_{\bm{\ell}'_2}\alpha_{\bm{L}})E_{\bm{\ell}_3}(} {E}
    \bcontraction {\langle E_{\bm{\ell}_1} (E_{\bm{\ell}'_2}}{\alpha}{_{\bm{L}}) E_{\bm{\ell}_3} (E_{\bm{\ell}'_4}}{\alpha}
    \langle E_{\bm{\ell}_1} (E_{\bm{\ell}'_2}\alpha_{\bm{L}}) E_{\bm{\ell}_3} (E_{\bm{\ell}'_4}\alpha_{\bm{L}'})\rangle} \rightarrow \mathrm{(b1)} \\
  + & {\contraction {\langle}{E}{_{\bm{\ell}_1} (E_{\bm{\ell}'_2}\alpha_{\bm{L}}) E_{\bm{\ell}_3}(}{E}
    \contraction [2ex] {\langle E_{\bm{\ell}_1}(} {E}{_{\bm{\ell}'_2}\alpha_{\bm{L}})} {E}
    \bcontraction {\langle E_{\bm{\ell}_1} (E_{\bm{\ell}'_2}}{\alpha}{_{\bm{L}}) E_{\bm{\ell}_3} (E_{\bm{\ell}'_4}}{\alpha}
    \langle E_{\bm{\ell}_1} (E_{\bm{\ell}'_2}\alpha_{\bm{L}}) E_{\bm{\ell}_3} (E_{\bm{\ell}'_4}\alpha_{\bm{L}'})\rangle} \rightarrow \mathrm{(b2)} \\
  + & {\contraction {\langle}{E}{_{\bm{\ell}_1(}}  {E}
    \contraction {\langle E_{\bm{\ell}_1}({E}{_{\bm{\ell}'_2}\alpha_{\bm{L}})}} {E}{_{\bm{\ell}_3}(}{E}
    \bcontraction {\langle E_{\bm{\ell}_1} (E_{\bm{\ell}'_2}}{\alpha}{_{\bm{L}}) E_{\bm{\ell}_3} (E_{\bm{\ell}'_4}}{\alpha}
  \langle E_{\bm{\ell}_1} (E_{\bm{\ell}'_2}\alpha_{\bm{L}}) E_{\bm{\ell}_3}  (E_{\bm{\ell}'_4}\alpha_{\bm{L}'})\rangle} \rightarrow \mathrm{(b3)},
  \end{split}
\end{equation}
where $(\mathrm{b1})$ is a disconnected term, and $(\mathrm{b2})$ and $(\mathrm{b3})$ are connected terms.

Due to the analogous contractions contributed by $\phi$
and $\alpha$ in the four-point function,
we can simply substitute $C_L^{\phi \phi}$
and $f^{\phi}$ with $C_L^{\alpha \alpha}$ and $f^{\alpha}$
in Eq.~\eqref{eq:clpp term} to express all
$\mathcal{O}(\alpha^{2})$ contribution. Applying
the orthogonality relation
Eq.~\eqref{eq:orthogonality relation}, $(\mathrm{b1})$ contributes an zero value as
\begin{equation}
  \label{eq:N1aa disconnected}
        \begin{aligned}
          C_L^{\alpha \alpha}\left(\frac{A_{\bm{L}}}{L^2}\right)^2 &\int \frac{d^2 \bm{\ell}}{(2 \pi)^2} \frac{f^{\phi}_{\bm{\ell}, \bm{L}-\bm{\ell}} f_{\bm{\ell}, \bm{L}-\bm{\ell}}^{\alpha} }{{\hat{C}^{\m{EE}}_{\ell} \hat{C}^{\m{BB}}_{|\bm{L}-\bm{\ell}|}}}\\
         \times &\int \frac{d^2 \bm{\ell}'}{(2 \pi)^2} \frac{f^{\phi}_{\bm{\ell}', \bm{L}-\bm{\ell}'} f_{\bm{\ell}', \bm{L}-\bm{\ell}'}^{\alpha}}{{\hat{C}^{\m{EE}}_{\ell'} \hat{C}^{\m{BB}}_{|\bm{L}-\bm{\ell}'|}}} = 0
      \end{aligned}
    \end{equation}
$(\mathrm{b2})$ and $(\mathrm{b3})$ generate a $N^{(1)}$-like bias named as $N^{(1, \alpha \alpha)}$ given by
\begin{equation}
      \begin{aligned}
        N^{(1, \alpha \alpha)}_{L} = &\frac{A^2_L}{L^4} \int \frac{d^2 \bm{\ell}}{(2\pi)^2} \frac{d^2 \bm{\ell'}}{(2\pi)^2} C^{\alpha \alpha}_{\bm{\ell'}} \frac{f^{\phi}_{\bm{\ell}, \bm{L}-\bm{\ell}}}{{\hat{C}^{\m{EE}}_{\ell} \hat{C}^{\m{BB}}_{|\bm{L}-\bm{\ell}|}}} \\
         & \times \frac{f^{\phi}_{\bm{\ell}-\bm{L}-\bm{\ell}', \bm{\ell}'-\bm{\ell}}}{{\hat{C}^{\m{EE}}_{|\bm{\ell}-\bm{L}-\bm{\ell}|} \hat{C}^{\m{BB}}_{|\bm{\ell}'-\bm{\ell}|}}}\ f^{\alpha}_{\bm{\ell},\bm{\ell}'-\bm{\ell}}f^{\alpha}_{\bm{L}-\bm{\ell}, \bm{\ell}-\bm{L}-\bm{\ell}'},
      \end{aligned}
    \end{equation}
    where the superscript $\alpha \alpha$ represents that this is
    at $\mathcal{O}(\alpha^2)$. We implement Eq.~\eqref{eq:N1aa}
    in the modified \texttt{plancklens} to calculate the
    analytic curves shown in Fig.~\ref{fig:N1aa_plancklens}.
    Eq.~\eqref{eq:N1aa sim} can then be obtained by simply
    substituting the lensed-only CMB maps $(\t{E}, \t{B})$
    in Eq.~\eqref{eq:N1 sim} by rotated-only ones $(E',B')$.

\bibliography{birefringence,lensing,cite}
\end{document}